\pdfoutput=1

\documentclass[runningheads]{llncs}
\usepackage{graphicx}
\usepackage{amssymb}
\usepackage{amsmath}
\usepackage{bm}
\usepackage{hyperref}
\usepackage{xcolor}
\usepackage{caption}
\usepackage{subcaption}
\usepackage{siunitx}
\usepackage{pifont}
\usepackage{lipsum}
\usepackage{float}
\usepackage{svg}
\usepackage{booktabs}

\begin{document}
%
% \title{Contribution Title\thanks{Supported by organization x.}}
\title{Fine-Tuned Self-Supervised Speech Representations for Language Diarization in Multilingual Code-Switched Speech}
\titlerunning{Language Diarization in Multilingual Code-Switched Speech}
\author{Geoffrey Frost\inst{1}\orcidID{0000-0002-6107-3858} \and Emily Morris\inst{2} \orcidID{0000-0003-0336-3903}
Joshua Jansen van Vüren\inst{1}\orcidID{0000-0002-3406-4788} \and
Thomas Niesler\inst{1}\orcidID{0000-0002-7341-1017}}
\authorrunning{G. Frost et al.}
\institute{Department of E\&E Engineering, Stellenbosch University, Stellenbosch, South Africa \inst{1}\\
\email{\{gfrost, jjvanvueren, trn\}@sun.ac.za}\inst{1}, \email{morrisemily0107@gmail.com}\inst{2}}

\maketitle              % typeset the header of the contribution
%
% either multi-lingual experts who can identify and transcribe these cases
\begin{abstract}
Annotating a multilingual code-switched corpus is a painstaking process requiring specialist linguistic expertise. This is partly due to the large number of language combinations that may appear within and across utterances, which might require several annotators with different linguistic expertise to consider an utterance sequentially. This is time-consuming and costly. It would be useful if the spoken languages in an utterance and the boundaries thereof were known before annotation commences, to allow segments to be assigned to the relevant language experts in parallel. To address this, we investigate the development of a continuous multilingual language diarizer using fine-tuned speech representations extracted from a large pre-trained self-supervised architecture (WavLM). We experiment with a code-switched corpus consisting of five South African languages (isiZulu, isiXhosa, Setswana, Sesotho and English) and show substantial diarization error rate improvements for language families, language groups, and individual languages over baseline systems.


\end{abstract}
\section{Introduction}

Prevalent in multilingual societies, code-switched (CS) speech is the phenomenon where two or more languages are used within the same conversation or utterance \cite{van2007grammar}. The development of automatic speech recognition (ASR) systems to model such speech typically require large amounts of data, which is often challenging to collect. Past attempts at compiling such a multilingual CS corpus have relied on a complex iterative process to transcribe audio recordings \cite{niesler2018first}. First, a principal transcriber segments a file into utterances and annotates their primary language. All unlabelled segments are passed to the next transcriber, who will in turn annotate their primary language. This iterative process is repeated until there are no more unlabelled segments. Prior knowledge of the specific language boundaries within a given utterance would allow this iterative process to be parallelised, saving time and reducing the human capital needed to orchestrate the previously described complex process. With the aim of extending a current multilingual CS corpus, we investigate the development of a simple end-to-end language diarization (LD) system to aid in this task.

There is limited recent literature that addresses LD for CS speech \cite{liu2021end}. Moreover, speech representations extracted from large pre-trained self-supervised acoustic models have yet to be leveraged in this domain despite being shown to perform well on a variety of downstream acoustic tasks \cite{yang2021superb}. In this work, we investigate the application WavLM \cite{chen2022wavlm}, a recent architecture of this kind that achieves state-of-the-art performance on a suite of down-stream language tasks, for LD \cite{yang2021superb}. We experiment with a corpus of low-resource multilingual code-switched soap opera speech, comprising English and four other low-resource South African languages (isiZulu, isiXhosa, Setswana, Sesotho). Although WavLM is pre-trained on monolingual English, we show that it transfers well to low-resource LD, and substantially improves upon previously proposed architectures for the same task.

\section{Background}

Broadly, both prelexical information (phonetic repertoire, phonotactics, rhythm and intonation) and lexical-semantic knowledge (meaning) is utilised when determining a spoken language \cite{ramus1999language}\cite{zhao2008cortical}, with inexperienced human listeners effectively able to identify languages relying only on the former, implying language identification (LID) can be performed with minimal content understanding \cite{muthusamy1994perceptual}\cite{van2008human}. Both phonotactic and acoustic features are effective representations when building systems to perform LID \cite{zissman1996comparison}\cite{muthusamy1994reviewing}.
% Automatic LID is an important prepossessing step in speech systems
Typically, phonotactic systems utilise a bank of monolingual large vocabulary continuous speech recognition (LVCSR) systems run in parallel, one for each language. The one producing the highest log-likelihood for the recognised word sequence is selected as the language spoken \cite{mendoza1996automatic}\cite{schultz1996lvcsr}\cite{hieronymus1996spoken}. This requires corpora with which to train each LVCSR system, which is not feasible for the low-resource languages often found in CS speech.

Unlike phonotactic approaches, acoustic systems attempt to learn a distribution across languages directly from acoustic features, and as such are more relevant to this work. Whilst early work relied on classical algorithms (GMMs, SVMs or LR) \cite{nakagawa1992speaker}\cite{yan1995development}\cite{van2006channel}\cite{brummer2010measuring}\cite{li2013spoken}, more recently, attention has shifted to end-to-end architectures using deep neural networks \cite{gonzalez2015frame}\cite{lopez2016use}\cite{trong2016deep}\cite{geng2016end}\cite{watanabe2017language}\cite{gelly2017spoken}\cite{cai2018insights}. However, these systems tend to solve a sequence-to-one problem, whereby a single embedding is extracted from a variable length utterance and hence assume only a single language is being spoken. This is insufficient for our task, for which spoken languages can change throughout a single utterance.

Whilst to the best of our knowledge there has been no work in LD for South African languages, recently one study has successfully demonstrated the use of end-to-end acoustic based systems for code-switched LD \cite{liu2021end}. In this approach, popular deep-learning-based speaker diarization approaches were applied to LD, yielding promising results. Long short-term memory networks and transformer-based systems were trained and evaluated on a corpus comprising $52$ hours of speech from three bilingual CS subcorpora (Gujarati-English, Tamil-English and Telugu-English). Systems trained for bilingual diarization and multilingual diarization performed well, with accuracies exceeding $80\%$.

\begin{figure}[h!]
\centering
\includegraphics[width=1\textwidth]{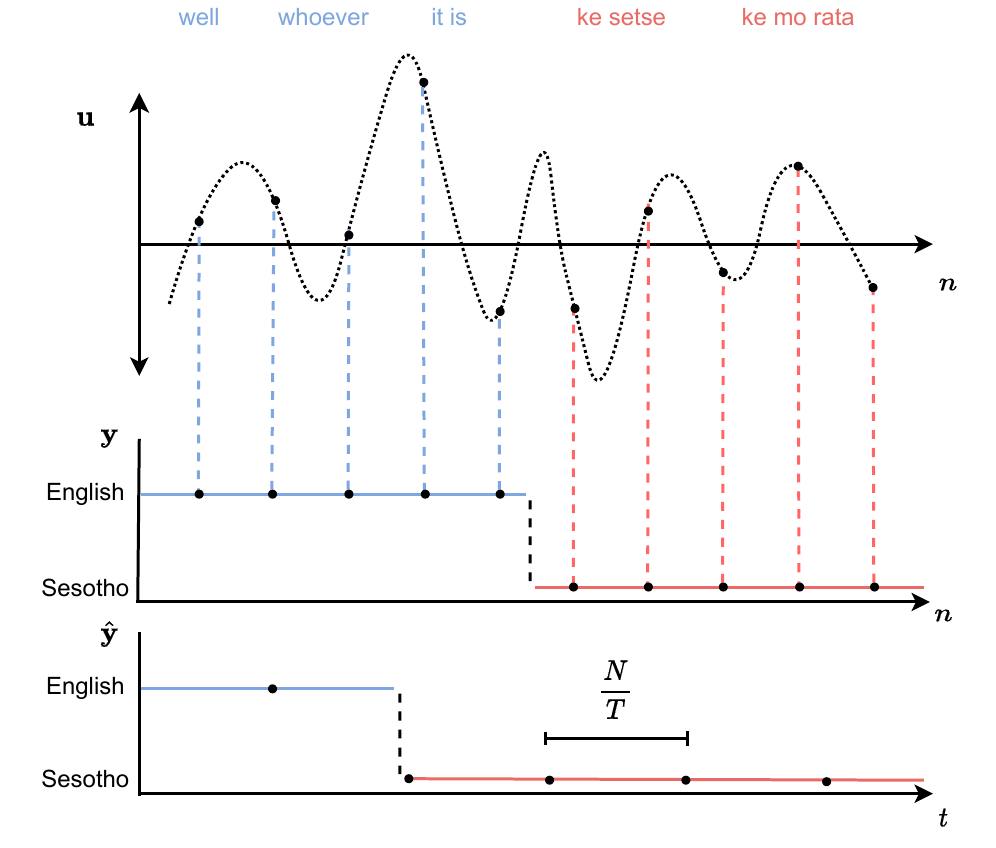}
\caption{A simple representation of a sampled code-switched utterance waveform $\mathbf{u}$ between English and Sesotho and the corresponding ground truth language labels $\mathbf{y}$, and those predicted by a LD system $\hat{\mathbf{y}}$.} \label{fig:ld_example}
\end{figure}

\vspace{-8mm}
\subsection{Language Diarization}

As it is fundamental to our work, we formally describe the task of LD, reinforced by a simple example. Given a sampled utterance waveform $\mathbf{u}=(u_{n}\in\mathbb{R}|1,...,N)$ with language labels $\mathbf{y}=(y_{n}\in[C]|n=1,...,N)$, where $N$ is the number of samples and $C$ is the set of languages that are to be identified, let the function $G(\mathbf{u};\theta)$, where $\theta$ are learnable parameters, be the function that estimates language labels $\hat{\mathbf{y}}=(\hat{y}_{t}\in[C]|1,...,T)$ for non-overlapping segments of $\mathbf{u}$ of sample length $\frac{N}{T}$, where $T$ is the number of segments. We illustrate this with simple CS example in Figure \ref{fig:ld_example}. Note how the period of predicted label segments is larger than the ground truth.

\section{Corpus}

We perform our LD experimentation using a multilingual CS corpus compiled from South African soap operas \cite{niesler2018first}. The corpus consists of 14.3 hours of annotated and segmented speech taken from 626 South African soap opera episodes, divided into four language-balanced subcorpora. Each subcorpus contains monolingual and CS utterances in English and one of four Bantu languages: isiZulu, isiXhosa, Setswana and Sesotho. We will refer to these subcorpora as English-isiZulu (E-Z), English-isiXhosa (E-X), English-Setswana (E-Se) and English-Sesotho (E-So). The subcorpora are each split into training, development, and test sets. The four Bantu languages in the corpus represent the two most widely spoken South African Bantu language groups. Namely, the Nguni languages (IsiZulu and isiXhosa), and the Sotho-Tswana languages (Setswana and Sesotho). Language groups are collections of languages with similar linguistic roots and characteristics. It is worth noting the proportional spread of the utterances across the subcorpora. Although each subcorpus is language-balanced, the E-Z subcorpus contains roughly twice as much data as the other subcorpora. The exact breakdown in terms of subcorpora is presented in Table \ref{tab:corpus_breakdown}.

The corpus contains utterances with a mixture of intersentential and intrasentential code switches, with the former occurring between sentences and the latter within sentences. Intrasentential code-switching can occur at a morpheme level within words, with a word such as \textit{amasponsors} being indicated as a switch from isiZulu to English. The rapid nature of the code-switching combined with the relatively fast pace of the soap opera speech requires high resolution LD which is hard to achieve \cite{niesler2018first}.

Even for soap opera speech, where language switches are frequent, many utterances remain monolingual. In total, only 6.3 of the 14.3 hours of speech correspond to utterances with code switches. Of this,  only $\approx$4.5 hours appear in the training set. However, while CS utterances are relatively sparse in the training set, the corpus design has ensured that they form a larger portion of the development set, whilst the test set is entirely comprised of CS speech. 

%For example, in the E-Z subcorpus, CS utterances make up roughly 36\% of the training set while the development and test sets are fully comprised of CS utterances.

\begin{table}[h!]
\centering
\small\addtolength{\tabcolsep}{5pt}
    \begin{tabular}{ c c c c c }
        \hline
        Subcorpus & Train & Dev & Test & Total \\ 
        \hline
        E-Z & 288.60 & 8.00 & 30.40 & 327.00\\ 
        E-X & 160.54 & 13.68 & 14.34 & 188.58\\
        E-Se & 139.74 & 13.83 & 17.83 & 171.60\\
        E-So & 141.72 & 12.77 & 15.54 & 169.80\\
        \hline
    \end{tabular}
    \vspace*{5mm}
    \caption{The amount of data (in minutes) in the training (train), development (dev) and test sets of the four subcorpora.}
    \label{tab:corpus_breakdown}
\end{table}

\section{Models}

We consider the application of three architectures for end-to-end LD. The first two approaches, which serve as our baselines, respectively utilise a two-stage bidirectional long short-term memory (BiLSTM) network and an x-vector transformer architecture as proposed in \cite{liu2021end}. We also introduce the pre-trained self-supervised acoustic representation model considered in this work, WavLM.

\subsection{BiLSTM}

\begin{figure}[tb!]
\centering
\includegraphics[width=1\textwidth]{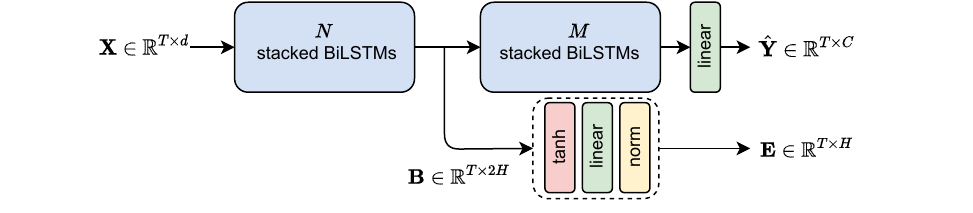}
\caption{Network diagram for the two-stage baseline BiLSTM LD system.} \label{fig:e2e_bilstm}
\end{figure}

Initially proposed for speaker diarization \cite{fujita2019end}, a two-stage BiLSTM architecture has been shown to perform well for LD \cite{liu2021end}. A sequence of $T$ acoustic feature vectors $\mathbf{X}=(\mathbf{x}_t\in\mathbb{R}^{d}|t=1,...,T)$ are extracted from an utterance. A set of $N$ BiLSTM layers are used to generate language representations $\mathbf{B}=(\mathbf{b}_t\in\mathbb{R}^{2H}|t=1,...,T)$ from $\mathbf{X}$, followed by $M$ BiLSTM layers to estimate the the sequence of language labels $\hat{\mathbf{Y}}=(\mathbf{y}_t\in\mathbb{R}^{|C|}|t=1,...,T)$ where $C$ is the set of languages that are to be identified. 
%A multi-objective loss term is used as described in Equation \ref{eq:multi_objective_loss}, where $\alpha$ is a regularisation parameter and $\mathbf{e}_t\in\mathbb{R}^{H}$ is an embedding of $\mathbf{b}_t$ as described in Figure \ref{fig:e2e_bilstm}. 
In addition to computing the frame-wise cross-entropy (CE) loss between the ground truth labels $\mathbf{Y}=(y_t\in[C]|1,...,T)$ and $\hat{\mathbf{Y}}$, a deep-clustering (DC) loss \cite{hershey2016deep} is used to encourage $\mathbf{B}$ to be language discriminative as shown in Equation \ref{eq:multi_objective_loss} where $\mathbf{e}_t\in\mathbb{R}^{H}$ is an emending of $\mathbf{b}_t$ and $\alpha$ is a regularisation parameter. A high-level depiction of this architecture is presented in in Figure \ref{fig:e2e_bilstm}.

%with language labels $\mathbf{Y}=(y_t\int{[C]|1,...,T)$ where $C$ is the number of potential languages

\begin{equation}\label{eq:multi_objective_loss}
\mathcal{L} = \alpha L_{CE}(y_t, \hat{\mathbf{y}}_t) + (1-\alpha)L_{DC}(y_t, \mathbf{e}_t)
\end{equation}

We use the same architectural hyper-parameters as previous applications of this architecture to LD of CS speech, with $H$, $N$, $M$, and $\alpha$ set to 256, 2, 3, and 0.5 respectively \cite{liu2021end}.

\subsection{X-vector Self-Attention}
\begin{figure}[tb!]
\centering
\includegraphics[width=1\textwidth]{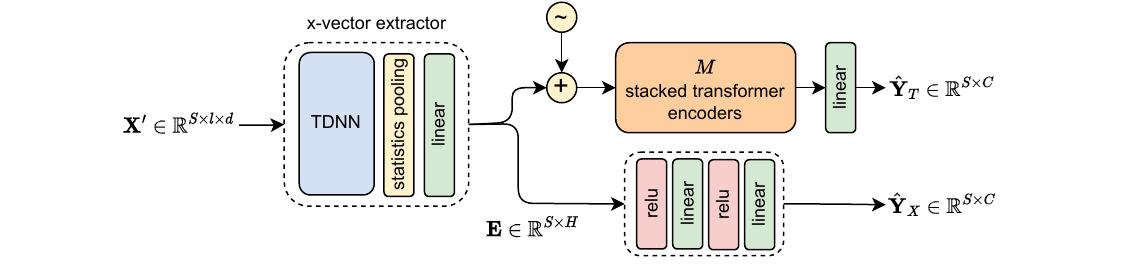}
\caption{Network diagram for the XSA LD system.} \label{fig:xsa}
\end{figure}

X-vector Self-Attention (XSA) is another end-to-end architecture proposed for LD in \cite{liu2021end}. First, x-vectors $\mathbf{E}=(\mathbf{e}_t\in\mathbb{R}^{H}|t=1,...,S)$ are extracted for non-overlapping segments of length $s$ of acoustic feature representations $\mathbf{X}^\prime=(\mathbf{x}_t\in\mathbb{R}^{s\times d}|t=1,...,S)$ for a given utterance. Note x-vectors simply refer to a fixed-sized neural embedding extracted for an arbitrary length of speech. $\mathbf{E}$ is then sinusoidally positionally encoded, and passed through a series of $M$ stacked self-attention (transformer encoder) modules, as illustrated in Figure \ref{fig:xsa}. The x-vector extractor's time-delayed neural network (TDNN) is made up of a series of temporal convolutional layers, the outputs of which are pooled over the time dimension (mean and variance) and linearly projected. The cross-entropy between $\mathbf{Y}=(y_t\in[C]|1,...,S)$ and segment level language predictions made with the output of the transformer network $\hat{\mathbf{Y}}_{T}=(\hat{\mathbf{y}}^{T}_t\in\mathbb{R}^{|C|}|t = 1,...,S)$, and the output of the x-vector extractor $\hat{\mathbf{Y}}_{X}=(\hat{\mathbf{y}}^{X}_t\in\mathbb{R}^{|C|}|t = 1,...,S)$ are used in a multi-objective loss as shown in Equation \ref{eq:multi_objective_loss_xsa} for a single segment, where $\alpha$ is a regularisation parameter. Whilst $\hat{\mathbf{Y}}_{X}$ is not used to make actual predictions come test time, its inclusion encourages the x-vector extractor to learn segment-level language information.

\begin{equation}\label{eq:multi_objective_loss_xsa}
\mathcal{L} = \alpha L_{CE}(y_t, \hat{\mathbf{y}}^{T}_t) + (1-\alpha)L_{CE}(y_t, \hat{\mathbf{y}}^{X}_t)
\end{equation}

We use the same architectural hyper-parameters as in previous applications of this architecture to LD of code-switched speech, with $H$, $M$, and $\alpha$ set to 256, 4, and 0.5 respectively \cite{liu2021end}.

\subsection{WavLM}

Large transformer-based acoustic language models have proven successful in capturing complex contextualised representations for a multitude of speech tasks \cite{yang2021superb}. WavLM \cite{chen2022wavlm} resembles other self-supervised speech representation frameworks such as wav2vec2.0 \cite{baevski2020wav2vec} and HuBERT \cite{hsu2021hubert}, whereby a temporal convolutional feature extractor extracts audio representations $\mathbf{X}=(\mathbf{x}_t\in\mathbb{R}^{H}|t=1,...,T)$ directly from a 16kHz waveform $\mathbf{u}$. These are subsequently fed into a large transformer encoder consisting of $L$ blocks, where the last encoder outputs a sequence of contextual representations $\mathbf{C}^{L}=(\mathbf{c}^{L}_t\in\mathbb{R}^{H}|t=1,...,T)$ which are used to solve a masked learning objective. Whilst other frameworks have achieved great success in speech recognition, they disregard information important for other speech tasks such as paralinguistics, speaker identity and semantics. WavLM addresses these shortcomings by introducing a masked speech denoising and prediction HuBERT-like loss term. For a given utterance $\mathbf{u}$, noisy or overlapped speech is manually simulated by sampling noise or other utterances from the batch to produce $\mathbf{u^{\prime}}$ which is then fed into the convolutional feature extractor generating the feature sequence $\mathbf{X}$ that corresponds to a 20ms framerate (i.e. a downsampling factor from waveform to feature representation of $320\times$). $M$ tokens in $\mathbf{X}$ are masked, and a set of discrete pseudo-labels $\mathbf{z}^{k}=(z^{k}_t\in[C^{k}]|t=1,...,T)$ are predicted from $\mathbf{C}^{L}$. The pseudo-labels are the acoustic unit cluster in the set of clusters $C^{k}$ that $\pmb{c}^{L}_t$ should belong to, discovered in an acoustic unit discovery step with either HuBERT embeddings (first-stage), or latent representations $\mathbf{C}^{L^\prime}$ extracted from the architecture itself (second-stage).
Predictions are made for $K$ acoustic unit sets, each with a different granularity (number of clusters) to facilitate the learning of different speech attributes. The distribution over the pseudo-labels $p$ is parameterized in the same way as HuBERT, resulting in the below loss function being used for self-supervised learning.

\begin{equation}\label{eq:hubert_loss}
\mathcal{L} = \sum_{k\in K}\sum_{t\in M}\log p^{k}(z^{k}_{t}|\mathbf{c}^{L}_{t})
\end{equation}

By keeping the original pseudo-labels for induced noisy/overlapped speech in an utterance, the network is forced to denoise the input, resulting in improved robustness for complex acoustic environments. Additionally, each transformer encoder consists of a convolution-based relative position embedding layer, which uses a relative position bias to allow the positional encoding to change based on the content of the input sequence \cite{chi2021xlm}.

\begin{figure}[tb!]
\centering
\includegraphics[width=1\textwidth]{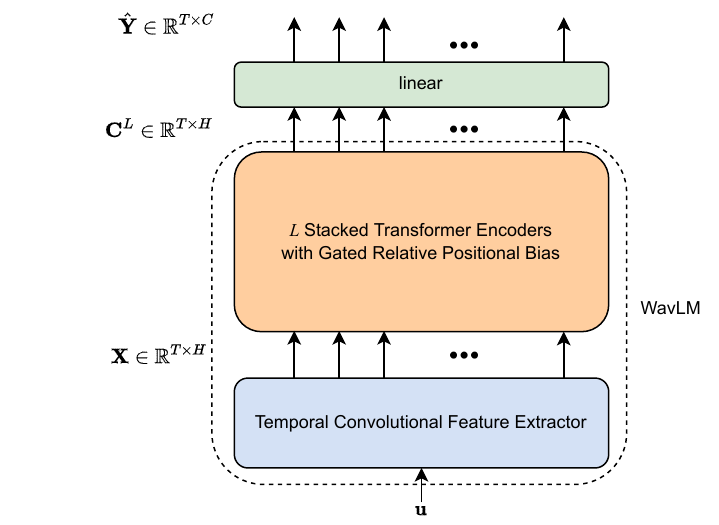}
\caption{Network diagram for the WavLM based LD system.} \label{fig:wavlm}
\end{figure}

WavLM is trained on an extremely large English corpus, which includes 94k hours from LibriLight \cite{kahn2020libri}, 10k hours from GigaSpeech \cite{chen2021gigaspeech} and 24k hours from VoxPopuli \cite{wang2021voxpopuli}. We use two of the released pre-trained models of varying size in our work, namely \textit{WavLM-base\texttt{+}} ($H=768$, $L=12$) and \textit{WavLM-large} ($H=1024$, $L=24$). For both, we use the contextual embeddings from the last transformer encoder layer $\mathbf{C}^{L}$ for language diarization by attaching a simple linear layer that maps each embedding $\mathbf{c}^{L}_{t}$ to a distribution over the possible $C$ languages being spoken over time $\hat{\mathbf{Y}}$. As with the BiLSTM baseline, we compute the frame-wise CE loss between $\hat{\mathbf{Y}}$ and $\mathbf{Y}$. A high-level depiction of the network is presented in Figure \ref{fig:wavlm}.

\section{Experimental Procedure}

To investigate the extent to which accurate diarization of code-switched speech can be achieved in the presence of multiple low-resource Bantu languages, each architecture is applied to three hierarchical diarization tasks:

\begin{enumerate}
  \item \textbf{English/Bantu}: All four Bantu languages are grouped and the network determines whether the language spoken in a segment is English or belongs to the Bantu family.
  \item \textbf{English/Nguni/Sotho-Tswana}: The Bantu languages are grouped according to their respective language groups and the network determines whether the language spoken in a segment is English, a Nguni language or a Sotho-Tswana language.
  \item\textbf{English/isiZulu/isiXhosa/Setswana/Sesotho}: The network determines whether the language spoken in a segment is English, isiZulu, isiXhosa, Setswana or Sesotho.
\end{enumerate}

Given the limited amount of training data and the substantial linguistic similarities between these languages, differentiating between various Bantu languages is a challenging task. In turn, the aforementioned three hierarchical tasks can be seen as increasing in complexity as we increase the number of Bantu language categories to be identified. Furthermore, while the monolingual utterances in the training data will still allow the architectures to learn underlying representations of the languages, the lack of CS utterances could affect their ability to correctly categorise segments in the presence of rapid language changes.

Both baseline systems are trained with the same configurations used in \cite{liu2021end}, with the exception that we increase the batch size to 64. When training the WavLM models, we use a learning rate of $\num{1e-4}$ and a weight decay of $\num{1e-4}$ with the AdamW optimizer. We use a batch size of 4 with 16 gradient accumulation steps, and train for 16 epochs. The learning rate is increased linearly for the first 1000 steps, followed by an exponential decay. We include label smoothing for all cross-entropy loss terms, set to $0.1$.  When training systems for tasks 2 and 3, we initialise model parameters with the corresponding architectures weights attained from training for the previous task, as it always resulted in improved performance during development. 
%That is, when training systems for task 2, we fine-tune the trained system for task 1 to task 2, and when training systems for task 3, we fine-tune the trained system for task 2 to task 3.
Additionally, for tasks 2 and 3 we remove all monolingual English utterances from the training set in an attempt to increase the proportional Bantu language representation across the set. All optimization and hyper-parameter tuning was conducted using the development set whilst we present and evaluate the performance of the architectures using the test set.  We make the source code used for experimentation available \footnote{\href{https://github.com/GeoffreyFrost/code-switched-language-diarization}{https://github.com/GeoffreyFrost/code-switched-language-diarization}}. %\href{https://github.com/GeoffreyFrost/code-switched-language-diarization}{here}.

\subsection{Data Preparation and Feature Extraction}

Samples in each mini-batch are zero-padded to the longest sequence in the respective batch and predictions made corresponding to these padded regions are disregarded during loss computation and evaluation. For transformer-based architectures, a padding mask is used to ignore these regions when computing self-attention. To acquire language labels for the segment-level language predictions made by each architecture, we first convert the time-stamped language boundaries provided by the corpus to a set of continuous language labels (a label for each sample of the utterance waveform) $\mathbf{y}=(y_n\in[C]|n=1,...,N)$, where $N$ is the number of samples in the digital waveform. $\mathbf{y}$ is then down-sampled to $\tilde{\mathbf{y}}=(\tilde{y}_t\in[C]|t=1,...,T)$, where $T$ is the number of segment predictions made by each architecture.

For both baseline systems we use 23-dimensional mel-spectrograms as acoustic feature vectors, with a frame length of $25\si{ms}$ and hop length of $10\si{ms}$ as in \cite{liu2021end}. Since the corpus we use provides language boundaries for each utterance and not labels for discrete segments, we do not have to further divide mel-spectrograms into $200\si{ms}$ (19 frames) segments as in \cite{liu2021end} for the BiLSTM architecture. However, we do have to conduct this division for the XSA architecture as the x-vector extractor is specifically designed to extract representations for such segments. In this case, the corresponding language label for each segment is the frame-wise language label that occurs the most often.

\subsection{Evaluation Metrics}

We use the error rate as our primary evaluation metric, which quantifies the proportion of incorrectly identified language segments as shown in Equation \ref{eq:error_rate}. Although a good representation of general system performance, the global error rate (GER), which computes the proportion of incorrect predictions across the entire evaluation set, can potentially be dominated by longer utterances. To quantify error rates on a  per-utterance level, we also compute the mean error rate (MER) across utterances.

\begin{equation}\label{eq:error_rate}
ER = \frac{Incorrect \; Predictions}{Total \; Predictions}
\end{equation}

\vspace{2pt}

\section{Results and Discussion}

 We present the error rates for the three diarization tasks using our chosen architectures in Table \ref{tab:test_results}. The two variations of the WavLM model achieve substantial improvements over the baseline architectures across all three tasks. The WavLM-large architecture provides the best overall performance, with the lowest GER and MER achieved being for task 1 ($10.05\%$ and $11.94\%$ respectively). The performance of all the architectures degrades with an increase in the granularity of the language categories which also results in a decrease in the number of training utterances per category. This is particularly prevalent in our two baseline models, with absolute GER increases of 14.72\% from task 1 to task 2 and by 10.54\% from task 2 to task 3 for the BiLSTM, and absolute increases of 12.4\% from task 1 to task 2 and by 10.80\% from task 2 to task 3 for the XSA architecture. These baseline architectures use randomly initialised weights and do not benefit from the same pre-training scheme as the WavLM architectures, making them more reliant on a larger amount of training data to achieve good results. However, the WavLM architectures do see a larger relative decrease in performance between task 2 and task 3. We also note that for all networks and tasks the GER is lower than the MER indicating that diarization error is dependent on the length of the utterance. This makes sense, as longer segments of speech have more contextual language information, which makes the language diarization task easier to perform.
 
\begin{table}[tb!]
\centering
\small\addtolength{\tabcolsep}{5pt}
    \begin{tabular}{ l c cc cc cc }
        \hline
                                 & $\#$Params & \multicolumn{2}{c}{Task 1}&\multicolumn{2}{c}{Task 2}&\multicolumn{2}{c}{Task 3} \\ 
                                 \cmidrule(lr){3-4}\cmidrule(lr){5-6}\cmidrule(lr){7-8}
                                 & & GER & MER & GER & MER & GER & MER \\
        \hline
        BiLSTM                   & 9M & 32.50         & 33.07          & 47.22         & 47.97          & 57.76          & 58.76\\
        XSA                      & 12M & 36.50         & 38.17          & 48.90         & 51.73          & 59.70          & 62.82\\
        WavLM-base\texttt{+}     & 95M & 12.57         & 14.87          & 15.96         & 19.17          & 33.55          & 37.18 \\
        WavLM-large              & 317M & \textbf{10.05}& \textbf{11.94} & \textbf{12.93}& \textbf{16.12} & \textbf{32.80} & \textbf{36.76}\\
        \hline
    \end{tabular}
        \vspace*{5mm}
     \caption{Test set GER and MER (\%) for each respective diarization task. \textit{Task 1} denotes \textit{English/Bantu} diarization, \textit{Task 2} diarization denotes \textit{English/Nguni/Sotho-Tswana}, and \textit{Task 3} denotes diarization of all languages.}
    \label{tab:test_results}
\end{table}

\begin{figure}[h!]
     \centering
     \begin{subfigure}[b]{0.49\textwidth}
         \centering
         \includegraphics[width=\textwidth]{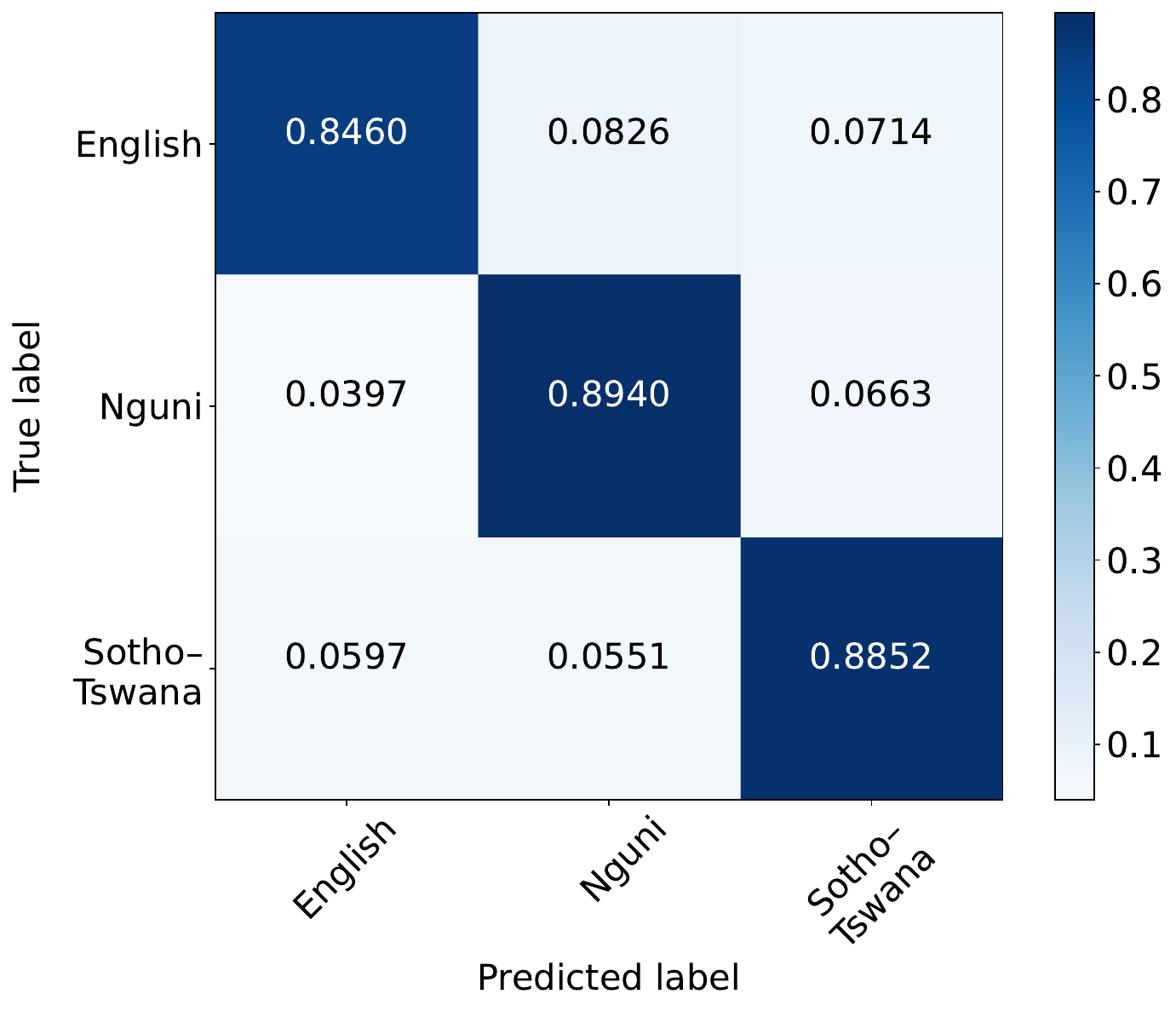}
         \caption{Task 2 confusion matrix.}
         \label{subfig:confusion_ENS}
     \end{subfigure}
     \hfill
     \begin{subfigure}[b]{0.49\textwidth}
         \centering
         \includegraphics[width=\textwidth]{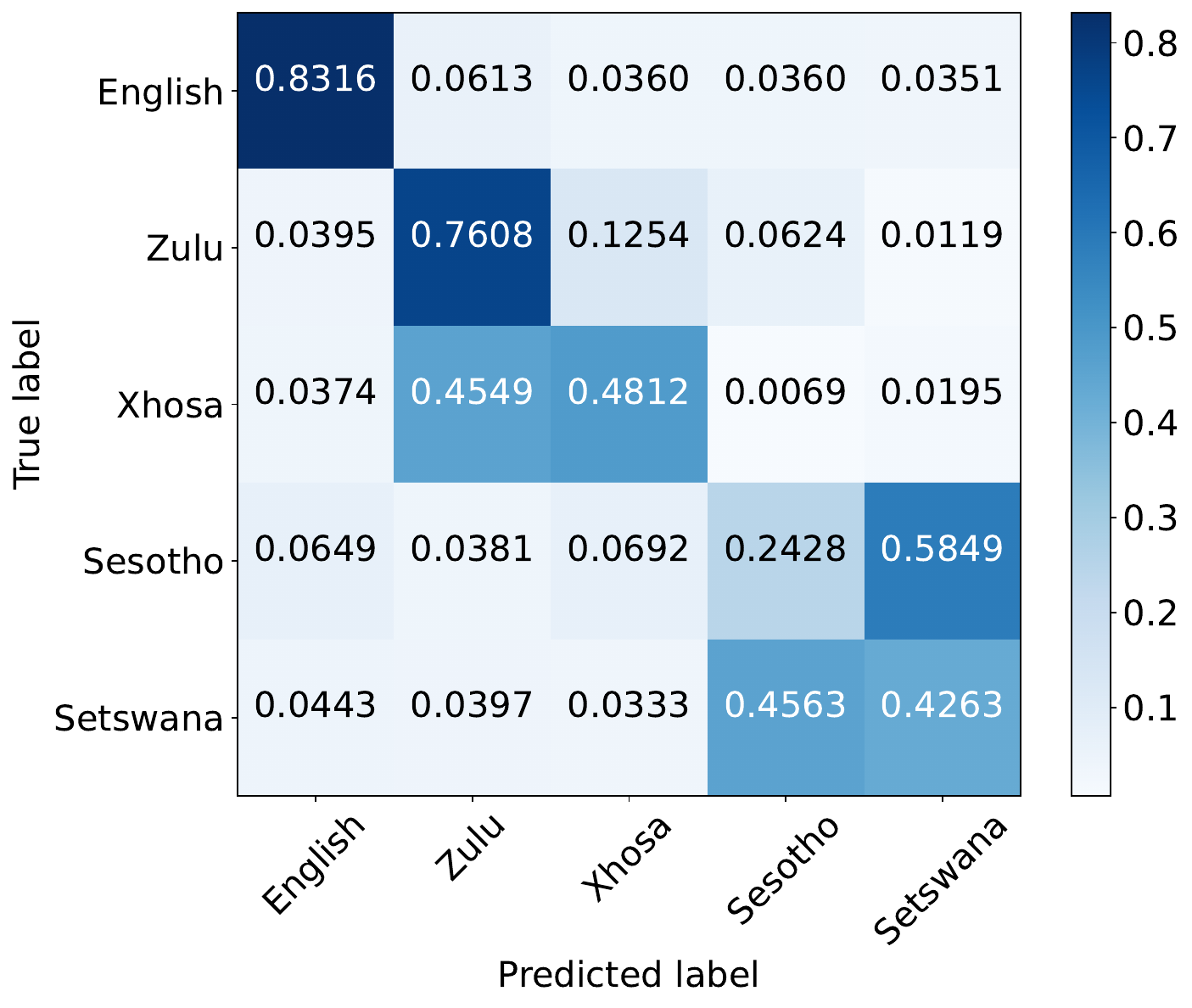}
         \caption{Task 3 confusion matrix.}
         \label{subfig:confusion_all}
     \end{subfigure}
     \caption{Confusion matrices depicting the accuracy of the WavLM-large architecture on Task 2 (English/Nguni/Sotho-Tswana) and Task 3 (all languages).}
     \label{fig:confusion_matrices}
\end{figure}

To further analyse the performance of WavLM and investigate the potential cause of the large performance degradation from task 2 to 3, we present the confusion matrices for both with predictions generated by the WavLM-large architecture in Figure \ref{fig:confusion_matrices}.
By analysing the confusion matrix for task 3 (Figure \ref{subfig:confusion_all}), its clear the degradation in performance is a result of incorrect language identification within the Nguni and Sotho-Tswana language groups. Albeit that WavLM's acoustic representations are comparatively effective for LD, these are grounded in English through its monolingual self-supervised pre-training scheme. Thus, the language structure that distinguishes languages within the same group, especially aspects that could not be learnt during English pre-training (e.g. syntax and phonology), are harder to learn.

Comparing the presented confusion matrices provides further insight into the influence of language groups on LD performance. Despite there being more training data available for the Nguni language group than for the Sotho-Tswana language group, by observing the confusion matrix presented in \ref{subfig:confusion_ENS} it is clear that accuracies for the two classes are essentially the same for task 2. The confusion matrix for task 3 presented in Figure \ref{subfig:confusion_all} shows how the Nguni languages benefit from this additional data, with improved isiZulu and isiXhosa accuracy compared to Sesotho and Setswana. However, as already discussed, there is a large degree of confusion within both Bantu language groups. Clearly the distinct nuances that distinguish languages within the same group are substantially more difficult to discern compared to those that differentiate groups within the same family. In addition to previously described effects of WavLM's monolingual English pre-training, this behaviour is potentially exacerbated by the use of the weights attained from task 2 to initialise the model for task 3, although the subsequent training should have tuned the model to correctly differentiate between the two classes within the Bantu and Sotho-Tswana language groups.

WavLM-large performs particularly well on identifying isiZulu, potentially due to the language being over-represented during training compared to other Bantu languages. This is further reinforced when considering the amount of isiXhosa misidentified as isiZulu, noting that for this language there is roughly half the amount of training data. In contrast, the equal (and lower) representation of both Sesotho and Setswana during training results in similar (and higher) confusion between the two.

\section{Conclusion}

In this work, we investigated the application of fine-tuned speech representations extracted from a large pre-trained self-supervised architecture (WavLM) for language diarization of code-switched speech. Through experimentation conducted with a code-switched corpus comprising five South African languages, we showed that utilising such an architecture can improve upon previously proposed systems for the same task. Despite being pre-trained on a monolingual corpus (English), WavLM was able to improve upon baseline systems when tasked with diarizing English/Bantu, English/Nguni/Sotho-Tswana and English/isiZulu/isiXhosa/Setswana/Sesotho coded-switched speech, reducing error rates by between $21.13\%$ and $31.85\%$ absolute compared to the best-performing baseline system. Whilst individual language accuracies are too low to aid in fully parallelised corpora annotation, surprisingly good performance was observed when performing language group diarization. Such performance may be sufficient to assign segments of utterances to language group streams (i.e. English, Nguni and Sotho-Tswana) reducing the number of language experts a segment may need to be sequentially reviewed by. 

\subsection{Limitations and Future Work} 
Due to computation and time constraints, this study was limited to exploring only one self-supervised architecture for LD. Additionally, disproportionate amounts of training data for each language may have influenced results. In future work, we aim to more rigorously investigate the application of self-surprised models to LD. This includes investigating if monolingual pre-training hinders the ability to reliably learn the discrete differences between languages within the same group by comparing the use of multilingual self-supervised models for the same task (such as wav2vec2-XLSR). 

\section{Acknowledgements}
This version of the article has been accepted for publication, after peer review and is subject to Springer Nature’s \href{https://www.springernature.com/gp/open-research/policies/accepted-manuscript-terms}{AM terms of use}, but is not the Version of Record and does not reflect post-acceptance improvements, or any corrections. The Version of Record is available online at: \href{https://doi.org/10.1007/978-3-031-22321-1_17}{https://doi.org/10.1007/978-3-031-22321-1\_17}
\bibliographystyle{splncs04}
\bibliography{paper}

\begin{thebibliography}{10}
\providecommand{\url}[1]{\texttt{#1}}
\providecommand{\urlprefix}{URL }
\providecommand{\doi}[1]{https://doi.org/#1}

\bibitem{baevski2020wav2vec}
Baevski, A., Zhou, Y., Mohamed, A., Auli, M.: wav2vec 2.0: A framework for
  self-supervised learning of speech representations. Advances in Neural
  Information Processing Systems  \textbf{33},  12449--12460 (2020)

\bibitem{brummer2010measuring}
Brummer, N.: Measuring, refining and calibrating speaker and language
  information extracted from speech. Ph.D. thesis, Stellenbosch: University of
  Stellenbosch (2010)

\bibitem{cai2018insights}
Cai, W., Cai, Z., Liu, W., Wang, X., Li, M.: Insights in-to-end learning scheme
  for language identification. In: Proceedings of IEEE International Conference
  on Acoustics, Speech and Signal Processing (ICASSP). pp. 5209--5213 (2018)

\bibitem{chen2021gigaspeech}
Chen, G., Chai, S., Wang, G., Du, J., Zhang, W.Q., Weng, C., Su, D., Povey, D.,
  Trmal, J., Zhang, J., et~al.: Gigaspeech: An evolving, multi-domain asr
  corpus with 10,000 hours of transcribed audio. In: Proceedings of Interspeech
  (2021)

\bibitem{chen2022wavlm}
Chen, S., Wang, C., Chen, Z., Wu, Y., Liu, S., Chen, Z., Li, J., Kanda, N.,
  Yoshioka, T., Xiao, X., et~al.: Wavlm: Large-scale self-supervised
  pre-training for full stack speech processing. IEEE Journal of Selected
  Topics in Signal Processing  (2022)

\bibitem{chi2021xlm}
Chi, Z., Huang, S., Dong, L., Ma, S., Singhal, S., Bajaj, P., Song, X., Wei,
  F.: Xlm-e: Cross-lingual language model pre-training via electra. arXiv
  preprint arXiv:2106.16138  (2021)

\bibitem{fujita2019end}
Fujita, Y., Kanda, N., Horiguchi, S., Nagamatsu, K., Watanabe, S.: End-to-end
  neural speaker diarization with permutation-free objectives. In: Proceedings
  of Interspeech (2019)

\bibitem{gelly2017spoken}
Gelly, G., Gauvain, J.L.: Spoken language identification using lstm-based
  angular proximity. In: Proceedings of Interspeech. pp. 2566--2570 (2017)

\bibitem{geng2016end}
Geng, W., Wang, W., Zhao, Y., Cai, X., Xu, B., Xinyuan, C., et~al.: End-to-end
  language identification using attention-based recurrent neural networks. In:
  Proceedings of Interspeech. pp. 2944--2948 (2016)

\bibitem{gonzalez2015frame}
Gonzalez-Dominguez, J., Lopez-Moreno, I., Moreno, P.J., Gonzalez-Rodriguez, J.:
  Frame-by-frame language identification in short utterances using deep neural
  networks. Neural Networks  \textbf{64},  49--58 (2015)

\bibitem{hershey2016deep}
Hershey, J.R., Chen, Z., Le~Roux, J., Watanabe, S.: Deep clustering:
  Discriminative embeddings for segmentation and separation. In: Proceedings of
  IEEE international conference on acoustics, speech and signal processing
  (ICASSP). pp. 31--35. IEEE (2016)

\bibitem{hieronymus1996spoken}
Hieronymus, J.L., Kadambe, S.: Spoken language identification using large
  vocabulary speech recognition. In: Proceedings of Fourth International
  Conference on Spoken Language Processing (ICSLP). pp. 1780--1783 (1996)

\bibitem{hsu2021hubert}
Hsu, W.N., Bolte, B., Tsai, Y.H.H., Lakhotia, K., Salakhutdinov, R., Mohamed,
  A.: Hubert: Self-supervised speech representation learning by masked
  prediction of hidden units. IEEE/ACM Transactions on Audio, Speech, and
  Language Processing  \textbf{29},  3451--3460 (2021)

\bibitem{kahn2020libri}
Kahn, J., Rivi{\`e}re, M., Zheng, W., Kharitonov, E., Xu, Q., Mazar{\'e}, P.E.,
  Karadayi, J., Liptchinsky, V., Collobert, R., Fuegen, C., et~al.:
  Libri-light: A benchmark for asr with limited or no supervision. In:
  Proceedings of IEEE International Conference on Acoustics, Speech and Signal
  Processing (ICASSP). pp. 7669--7673. IEEE (2020)

\bibitem{li2013spoken}
Li, H., Ma, B., Lee, K.A.: Spoken language recognition: from fundamentals to
  practice. Proceedings of the IEEE  \textbf{101}(5),  1136--1159 (2013)

\bibitem{liu2021end}
Liu, H., Garc{\'\i}a-Perera, L.P., Zhang, X., Dauwels, J., Khong, A.W.,
  Khudanpur, S., Styles, S.J.: End-to-end language diarization for bilingual
  code-switching speech. In: Proceedings of Interspeech. pp. 1489--1493 (2021)

\bibitem{lopez2016use}
Lopez-Moreno, I., Gonzalez-Dominguez, J., Martinez, D., Plchot, O.,
  Gonzalez-Rodriguez, J., Moreno, P.J.: On the use of deep feedforward neural
  networks for automatic language identification. Computer Speech \& Language
  \textbf{40},  46--59 (2016)

\bibitem{mendoza1996automatic}
Mendoza, S., Gillick, L., Ito, Y., Lowe, S., Newman, M.: Automatic language
  identification using large vocabulary continuous speech recognition. In:
  Proceedings of IEEE International Conference on Acoustics, Speech, and Signal
  Processing (ICASSP). pp. 785--788 (1996)

\bibitem{muthusamy1994reviewing}
Muthusamy, Y.K., Barnard, E., Cole, R.A.: Reviewing automatic language
  identification. IEEE Signal Processing Magazine  \textbf{11}(4),  33--41
  (1994)

\bibitem{muthusamy1994perceptual}
Muthusamy, Y.K., Jain, N., Cole, R.A.: Perceptual benchmarks for automatic
  language identification. In: Proceedings of IEEE International Conference on
  Acoustics, Speech and Signal Processing (ICASSP). pp. I--333 (1994)

\bibitem{nakagawa1992speaker}
Nakagawa, S., Ueda, Y., Seino, T.: Speaker-independent, text-independent
  language identification by hmm. In: Proceedings of Second International
  Conference on Spoken Language Processing (1992)

\bibitem{ramus1999language}
Ramus, F., Mehler, J.: Language identification with suprasegmental cues: A
  study based on speech resynthesis. The Journal of the Acoustical Society of
  America  \textbf{105}(1),  512--521 (1999)

\bibitem{schultz1996lvcsr}
Schultz, T., Rogina, I., Waibel, A.: Lvcsr-based language identification. In:
  Proceedings of IEEE International Conference on Acoustics, Speech, and Signal
  Processing (ICASSP). pp. 781--784 (1996)

\bibitem{trong2016deep}
Trong, T.N., Hautam{\"a}ki, V., Lee, K.A.: Deep language: a comprehensive deep
  learning approach to end-to-end language recognition. In: Proceedings of
  Odyssey: The Speaker and Language Recognition Workshop. vol.~2016, pp.
  109--116 (2016)

\bibitem{van2007grammar}
Van~Dulm, O.: The grammar of English-Afrikaans code switching: A feature
  checking account. Ph.D. thesis, External Organizations (2007)

\bibitem{van2006channel}
Van~Leeuwen, D.A., Brummer, N.: Channel-dependent gmm and multi-class logistic
  regression models for language recognition. In: Proceedings of Odyssey: The
  Speaker and Language Recognition Workshop. pp.~1--8 (2006)

\bibitem{van2008human}
Van~Leeuwen, D.A., De~Boer, M., Orr, R.: A human benchmark for the nist
  language recognition evaluation 2005. In: Proceedings of Odyssey: The Speaker
  and Language Recognition Workshop. p.~12 (2008)

\bibitem{wang2021voxpopuli}
Wang, C., Riviere, M., Lee, A., Wu, A., Talnikar, C., Haziza, D., Williamson,
  M., Pino, J., Dupoux, E.: Voxpopuli: A large-scale multilingual speech corpus
  for representation learning, semi-supervised learning and interpretation.
  arXiv preprint arXiv:2101.00390  (2021)

\bibitem{watanabe2017language}
Watanabe, S., Hori, T., Hershey, J.R.: Language independent end-to-end
  architecture for joint language identification and speech recognition. In:
  Proceedings of IEEE Automatic Speech Recognition and Understanding Workshop
  (ASRU). pp. 265--271 (2017)

\bibitem{niesler2018first}
van~der Westhuizen, E., Niesler, T.: A first south african corpus of
  multilingual code-switched soap opera speech. In: Proceedings of the Eleventh
  International Conference on Language Resources and Evaluation (LREC) (2018)

\bibitem{yan1995development}
Yan, Y.: Development of an approach to language identification based on
  language-dependent phone recognition. Oregon Graduate Institute of Science
  and Technology (1995)

\bibitem{yang2021superb}
Yang, S.w., Chi, P.H., Chuang, Y.S., Lai, C.I.J., Lakhotia, K., Lin, Y.Y., Liu,
  A.T., Shi, J., Chang, X., Lin, G.T., et~al.: Superb: Speech processing
  universal performance benchmark. In: Proceedings of Interspeech (2021)

\bibitem{zhao2008cortical}
Zhao, J., Shu, H., Zhang, L., Wang, X., Gong, Q., Li, P.: Cortical competition
  during language discrimination. NeuroImage  \textbf{43}(3),  624--633 (2008)

\bibitem{zissman1996comparison}
Zissman, M.A.: Comparison of four approaches to automatic language
  identification of telephone speech. IEEE Transactions on speech and audio
  processing  \textbf{4}(1), ~31 (1996)

\end{thebibliography}
\end{document}